 \documentstyle[epsf,12pt]{article}
 
\newcommand{\be}{\begin{equation}}
\newcommand{\ee}{\end{equation}}
\newcommand{\bea}{\begin{eqnarray}}
\newcommand{\eea}{\end{eqnarray}}
\newcommand{\beann}{\begin{eqnarray*}}
\newcommand{\eeann}{\end{eqnarray*}}
\newcommand{\beasn}{\begin{sneqnarray}}
\newcommand{\eeasn}{\end{sneqnarray}}

\newcommand{\al}{\alpha}

\newcommand{\Appendix}[1]%
    {\renewcommand{\thesection}{Appendix~\Alph{section}}%
     \section{#1}%
     \renewcommand{\thesection}{\Alph{section}} }

\def\bfnabla{\mbox{\boldmath $\nabla$}}

\def\bfsigma{\mbox{\boldmath $\sigma$}}
\newcommand{\nn}{\nonumber}

\catcode`@=11
\def\dif{{\rm d}}
\def\deriv{\@ifnextchar[{\@deriv}{\@deriv[]}}
   \def\@deriv[#1]#2#3{\mathchoice%
{{\dif^{#1}#2\over\dif{#3}^{#1}}}{{\dif^{#1}#2/\dif{#3}^{#1}}}%
{{\dif^{#1}#2\over\dif{#3}^{#1}}}{{\dif^{#1}#2/\dif{#3}^{#1}}}}

\def\presup#1{{}^{#1}\kern-.15em\relax}      
\def\presub#1{{}_{#1}\kern-.12em\relax}      

\def\secteqno{\@addtoreset{equation}{section}%
\def\theequation{\thesection.\arabic{equation}}}
\def\endsecteqno{\def\theequation{\@ifundefined{chapter}%
{\arabic{equation}}{\thechapter.\arabic{equation}}}}
\newcounter{subequation}
\def\thesubequation{\alph{subequation}}
\def\sneqnarray{\stepcounter{equation}\let\@currentlabel=\theequation
\setcounter{subequation}{1}
\def\@eqnnum{{\rm (\theequation\thesubequation)}}
\global\@eqcnt\z@\tabskip\@centering\let\\=\@eqncr\let\@@eqncr=\@@sneqncr
$$\halign to \displaywidth\bgroup\@eqnsel\hskip\@centering
 $\displaystyle\tabskip\z@{##}$&\global\@eqcnt\@ne
 \hskip 2\arraycolsep \hfil${##}$\hfil
 &\global\@eqcnt\tw@ \hskip 2\arraycolsep $\displaystyle\tabskip\z@{##}$\hfil
  \tabskip\@centering&\llap{##}\tabskip\z@\cr}
\def\endsneqnarray{\@@sneqncr\egroup $$\global\@ignoretrue}
\def\@@sneqncr{\let\@tempa\relax
   \ifcase\@eqcnt \def\@tempa{& & &}\or \def\@tempa{& &}
   \else \def\@tempa{&}\fi
     \@tempa \if@eqnsw\@eqnnum\stepcounter{subequation}\fi
     \global\@eqnswtrue\global\@eqcnt\z@\cr}
\def\nobiblabels{\def\@lbibitem[##1]##2{\@bibitem{##2}}}
\catcode`@=12

\secteqno
\begin{document}\setlength{\unitlength}{1mm}

\newcommand{\feynbox}[2]{\mbox{\parbox{#1}{#2}}}


\def\s{\sigma}
\def\g{\gamma}
\def\m{\mu}
\def\n{\nu}
\def\d{\delta}
\def\als{\alpha_{s}}


\title{{\bf Potential NRQED: The Positronium Case
      }}

\author{A. Pineda\\
       {\it \small Forschungszentrum J\"ulich, Institut f\"ur Kernphysik (Theorie)}\\
       {\it \small D-52425 J\"ulich, Germany.}\\
        \\
        and\\
        \\
        J. Soto\\
       {\it \small Dept. d'Estructura i Constituents de la Mat\`eria and IFAE}\\
       {\it \small Universitat de Barcelona. Diagonal, 647} \\
       {\it \small E-08028 Barcelona, Catalonia, Spain.}\\
        {\it e-mails:} \small{a.pineda@fz-juelich.de, soto@ecm.ub.es} }

\date{\today}

\maketitle

\thispagestyle{empty}

\begin{abstract}

We discuss in detail potential NRQED (pNRQED), a previously proposed
effective field theory for ultrasoft photons. The pNRQED lagrangian
for the equal mass case is presented and it is shown that it correctly
reproduces the positronium spectrum at order $m\alpha^5$. The pNRQED
lagrangian for the unequal mass case is also presented
at the same order. Dimensional regularization is used throughout.

\end{abstract}

\medskip

Keywords: Effective Field Theories, NRQED, NRQCD, HQET, Matching.

PACS: 11.10.St, 12.20.-m, 12.20.Ds, 12.39.Hg.

\vfill
\vbox{
\hfill May 1998\null\par
\hfill UB-ECM-PF 98/11\null\par
\hfill KFA-IKP(TH)-98-9}\null\par

\clearpage



\section{Introduction}
\indent

\bigskip


Non-relativistic QED (NRQED) \cite{Lepage} is becoming increasingly popular for 
QED bound state calculations \cite{Lepage,Kinoshita,Labelle2,LH}. It has the 
advantage over traditional Bethe-Salpeter
equations \cite{BS} that the non-relativistic nature of the
QED bound states is explicit, whereas relativistic and radiative corrections can be
systematically incorporated by taking into account higher orders in the $1/m$
expansion and by calculating the matching coefficients at higher order in $\alpha$
respectively. However, the NRQED lagrangian still contains two dynamical scales, namely
the typical relative momentum in the bound state ${\bf p}\sim m\alpha$ and the bound
state energy $E \sim m\alpha^2$, which implies that the terms in the 
lagrangian do not have a unique size. The leading size of each term is given by the
next relevant scale {\bf p} (soft) (except for the time derivative) 
and rules have been provided to
estimate the subleading contributions due to the scale $E$ (ultrasoft) \cite{Labelle}.
 Nevertheless,
it would be helpful for bound state calculations to have and effective field theory
(EFT)
where each term in the lagrangian had a well defined size. This EFT has proven to be 
 quite elusive for some time \cite{LM,Grin,Sav}. 

\medskip

In ref. \cite{Mont} we proposed potential NRQED (pNRQED) as such EFT and presented
the form of its lagrangian for positronium. In ref. \cite{Lamb} we worked out pNRQED
for Hydrogen-like atoms and reproduced the Lamb shift in a very straightforward way. 
It is the aim of this paper to discuss pNRQED for positronium in greater detail and to 
show that it also allows to reproduce the spectrum at order
$m\alpha^5$, where all 
regions of momenta (hard, soft and ultrasoft) contribute,
 very efficiently. 
 We also illustrate how dimensional regularization helps
in that.

\medskip

pNRQED describes fermion-antifermion pairs with relative 
momentum of order ${\bf p}$ and energy of order $E$, and
ultrasoft photons with energy and momentum of order $E$. This should 
be compared with NRQCD, which describes degrees of freedom (fermions and 
photons) with energy and momentum less than a certain cut-off $\mu$ such that 
$E,
{\bf p} << \mu << m$.
Formally speaking pNRQED has two UV cut-offs $\Lambda_1$ and
$\Lambda_2$, where $ E << \Lambda_1 << {\bf p}$ is the cut-off for the
energy of the fermions and for the energy and momentum of the ultrasoft photons,
whereas ${\bf p} << \Lambda_2 << m$ is the cut-off for the relative
momentum of the fermion-antifermion system. In principle, we have some
freedom to choose the relative importance between $\Lambda_1$ and
$\Lambda_2$. We choose $\Lambda_2^2/m <<  \Lambda_1$, which guarantees
that the UV behavior of the fermion propagators in pNRQED is that of the static ones.

pNRQED is obtained from NRQED by integrating out fermions and photons of energies 
and momenta of order ${\bf p}$ and photons of energies of order $E$ and 
momenta ${\bf p}$\footnote{Some authors \cite{Beneke,Harald} like to distinguish
 {\it potential} photons,
 i.e. photons with
$k^0\sim E$ and ${\bf k}\sim {\bf p}$ from {\it soft} photons, i.e. photons with
$k^0\sim {\bf p}$ and  ${\bf k}\sim {\bf p}$. This distinction is quite irrelevant in
our formulation since both {\it potential} and {\it soft} photons are integrated out
at the same time when matching NRQED to pNRQED.}.
The pNRQED lagrangian obtained is local in time but non-local in
space (i.e. it has potential terms) and contains ultrasoft photons only. 
The size 
of each
term becomes explicit once the
lagrangian is projected onto the one electron one positron subspace of the Fock space
and written in terms of a wave function field. This is due to the fact that in the 
latter representation the
ultrasoft photon fields can be multipole expanded about the center of mass.
Moreover, the calculations in pNRQED are very close to those in
nonrelativistic Quantum Mechanics.  
 
\medskip


The practical way in which we integrate out degrees of freedom is by a matching
procedure. We impose two fermion Green functions and four fermion Green functions
 (with an arbitrary number of
ultrasoft photon legs) in NRQED be equal to those in pNRQED once both are expanded about
the external fermion energies and (ultrasoft) photon energies and momenta. Dimensional
regularization is used for both UV and IR divergences. Furthermore, we use static
propagators, and hence the matching can be done to a given order in $1/m$ and $\alpha$.
This is justified because the fermion energies we are integrating out
in loops are of order
$m\alpha$ whereas the typical kinetic energy is
$O(m\alpha^2)$. 

\medskip

The procedure above is similar to the matching between QED and
NRQED as carried out in refs. \cite{Mont,Manohar,Match}. In that case scales
 $\sim m$  are integrated out and the matching reduces to
 calculations in QED, where $m$ is the only scale in the integrals. Here, the scales
which are integrated out are $\sim {\bf p}$ and the matching reduces to calculations in
NRQED where 
  ${\bf p}$  is the only scale in the integrals. Hence, the potential terms in pNRQED play a
r\^ole analogous to the Wilson coefficients in NRQED. Indeed, the former encode
contributions due to physics at the scale  ${\bf p}$ much in the same
way as the latter do  
of physics at the scale $m$. At each matching step the non analytic
behavior in the scale which is integrated out becomes explicit.

\medskip

We distribute
the paper as follows. In sec. 2 we describe the matching procedure between NRQED and
pNRQED. In sec. 3 we present the bound state calculation.
Sec. 4 is devoted to the conclusions and future prospects. In 
Appendix A gauge independence is checked at order $m\al^4$ by
calculating the matching to this order in the Feynman gauge. In
Appendix B the pNRQED lagrangian for the unequal mass
case is displayed.

\bigskip

\section{Matching NRQED to pNRQED}
\indent

\medskip

The pieces of the NRQED lagrangian which are relevant to the calculation of the bound
state energy at $O(m\alpha^5)$ read
\bea
\label{lagnrqed}
&&{\cal L}_{NRQED}= \psi^\dagger \Biggl\{ i D^0 + \, {{\bf D}^2\over 2 m} +
 {{\bf D}^4\over
8 m^3} + c_F\, e {{\bf \bfsigma \cdot B} \over 2 m}
 + c_D\, e {\left[{\bf \bfnabla \cdot E }\right] \over 8 m^2} \nonumber \\
&&
\quad\quad
 +
 i c_S\, e {{\bf \bfsigma \cdot \left(D \times E -E \times D\right) }\over 8 m^2}
\Biggr\} \psi
+
(\chi_c,\; e \rightarrow -e)
\\  
&&
\quad\quad
\nn
-
{d_{s} \over m^2} \psi^{\dag} \psi \chi_c^{\dag} \chi_c
+
  {d_{v} \over m^2} \psi^{\dag} {\bfsigma} \psi
                         \chi_c^{\dag} {\bfsigma} \chi_c
- {1\over 4} F_{\mu \nu} F^{\mu \nu}
+{d_2\over m^2} F_{\mu \nu} D^2 F^{\mu \nu}
\nonumber
\,,
\eea
where $\psi$ is the Pauli spinor field that annihilates the 
fermion and $\chi_c$\footnote{We use here $\chi_c=C\chi^{\ast}$, where $C$ is the charge conjugation
matrix, instead of $\chi$ as in refs. \cite{Mont,Lamb} because it is more usual
  in nonrelativistic systems.} is the Pauli spinor field that
annihilates the 
antifermion. $ i D^0=i\partial_0 -eA^0$, $i{\bf D}=i{\bfnabla}+e{\bf A}$
on the fermion field. The bilinear lagrangian for $\chi_c$ is equal to the
$\psi$ one with the change $e \rightarrow -e$. 

\medskip

This can be seen as follows. We draw all possible diagrams of the two fermion 
irreducible four fermion Green function, which can not be disconnected by cutting 
a photon line, such that $s+r\le 4$, where $s$ is the number of $1/m$ factors in 
the diagram and $r$ the number of explicit $\alpha$ \cite{Labelle,Mont,Lamb}. This 
rule can be easily justified if we take into account that the next relevant scale 
is ${\bf p} \sim m\alpha$ and hence all $1/m$ must be compensated for by $m\alpha$ 
until we reach dimensions of energy. For diagrams which can be disconnected by 
cutting a photon line the same rule applies but there is an extra suppression 
if $n$ time derivatives act on this photon line. This is due to the fact that these 
time derivatives are only sensitive to the typical energy. The extra suppression 
factor is $\al^n$. Recall
also that any diagram may have subleading contributions which appear from the analytic
expansion of the external energy about zero.
In the Coulomb gauge the diagrams fulfilling the above criteria are displayed in fig.
1 and fig. 2. Then both tree level and one loop diagrams are required. Notice also
that $c_{F}$, $c_{D}$ and $c_{S}$ are needed at order $\alpha$ and $d_{s}$ and $d_{v}$
at order $\alpha^2$. All these Wilson coefficients are gauge independent but depend on
the renormalization scheme for the UV divergences of NRQED. We shall use dimensional
regularization with $\overline{MS}$ scheme. In this scheme the Wilson coefficients of
the bilinear terms were given in \cite{Manohar} and the ones for the four fermion
terms in \cite{Mont,Match}. They read

\bea
\label{coefnrqed}
c_F &=& 1+{\alpha\over 2 \pi} \nn \,,\\
c_D &=& 1+{\alpha\over \pi} \left(
{4\over 3} \log {m^2 \over \mu^2} \right) \nn \,,\\
c_S &=& 1+{\alpha\over \pi} \,, \nn \\
d_2 &=& {\alpha\over 60 \pi} \,, \nn 
\\
d_{s} &=&
{3 \pi \al \over 2}
\Biggl\{ 1
  - {2 \al \over 3 \pi}
\left(  \log{m^2 \over  \mu^2}
                   + {23 \over 3} - \log2 + i {\pi \over 2} \right)
\Biggr\}
\,,
\\
\nn
d_{v} &=& 
-{ \pi \al \over 2}
\Biggl\{ 1
  - {2 \al \over  \pi}
\left(  {22 \over 9} + \log2 - i {\pi \over 2} \right)
\Biggr\}
\,.
\eea

\medskip

Hence our starting point is the lagrangian (\ref{lagnrqed}) with the Wilson 
coefficients (\ref{coefnrqed}).
We wish to integrate out fermions and photons of energy and momenta $\sim {\bf p}$ and 
photons of
energy $\sim E$ and momentum $\sim {\bf p}$. Then the effective theory we want to
reach, namely pNRQED,
 contains only
ultrasoft photons (of energy and momentum $\sim E$) and fermions
of energy $\sim E$ and momentum $\sim {\bf p}$ or less. Since we integrate out photon momenta 
$\sim {\bf p}$ but keep fermion momenta of this order, pNRQED contains terms non-local
in space, namely potential terms. 
 This is not a problem for a
non-relativistic EFT.

\medskip

The practical way to obtain pNRQED is by enforcing two and four fermion Green
functions with arbitrary ultrasoft external photons to be equal to those of NRQED 
once we expand about
zero
the energy in the external electron legs and the energy and momenta of the ultrasoft
photon legs. This may produce IR divergences which are regulated in dimensional regularization,
 in the same
way as the
UV divergences are. Since the IR behavior of NRQED and pNRQED is the same, these IR
divergences will cancel out in the matching. The UV divergences of NRQED must be
renormalized in the $\overline{MS}$ if we want to use the matching coefficients (\ref{coefnrqed}). We
still have a choice in the renormalization scheme of pNRQED. However, it is most
 advantageous to use again $\overline{MS}$. Indeed, in this scheme we can blindly use 
$\overline{MS}$ for any divergence regardless it is UV or IR in the matching
calculation. For the UV divergences of NRQED and pNRQED it is just the scheme we
 choose, and
for the IR divergences it is irrelevant as long as we use the same treatment in both
theories, since the IR behavior is the same. This allows to put integrals with no
scale equal to zero and, as we will see later on, reduces the matching to a calculation
 in NRQED only. 

\medskip

Notice that we demand off-shell Green functions in NRQED and pNRQED to
be equal and not on-shell Green functions as it is costume in the matching from QED to NRQED \cite{Lepage,Manohar}. This is due to the
fact that we are eventually interested in bound states, and particles in a bound state are typically
off-shell.
The equations of
motion of pNRQED (with potential terms included), or local field
redefinitions, may be consistently used later on to remove time derivatives in
higher order terms and write the pNRQED lagrangian in a standard form,
within the philosophy advocated in ref. \cite{Fearing} (see also
\cite{Balzereit}). We have checked in Appendix A that this procedure
produces gauge independent results at $O(m\alpha^4)$\footnote{However,
  there is still some freedom in the choice of the wave function
field, to be introduced later on, due to time independent unitary
transformations which commute with the leading terms in the pNRQED
lagrangian. Therefore, in general, it is not to be expected that the
standard form of the pNRQED lagrangian calculated with different
gauges coincide, but only to be related by one such unitary
transformation. This explains, for instance, why the potential
presented in \cite{positronium} is different
 from ours but leads to the same physics.}. In fact the same argument
applies to the matching between QED and NRQED, which accordingly
should also be carried out off-shell. However, in that case, at lower
orders there is no difference between doing the matching on-shell and
replacing derivatives by covariant derivatives, in order to enforce
gauge invariance, and doing the matching off-shell and consistently
using the equations of motion or local field redefinitions to get rid
of time derivative terms \cite{Manohar,Fearing}. 

\medskip

The remaining important ingredient to carry out the matching
efficiently is the use of static (HQET) propagators for the fermions. 
This has been 
completely justified in the matching between QED and NRQED
\cite{Manohar,Match}, since both energy and momentum in
this theory
cannot exceed the same cut-off $\Lambda$ which is smaller than $m$. Hence the UV 
behavior of the fermion
propagator in NRQED is always dominated by the energy. This fact is not
automatically implemented in dimensional regularization. When dimensional regularization is used,
 the correct UV behavior of NRQED is
only obtained when expanding about the static propagator.

\medskip

In pNRQED we have a certain choice for the UV cut-offs for the fermion energy and
momentum. We shall choose $\Lambda_1 << {\bf p}$ for the energy and $\Lambda_2 << m$ for
the momentum in such a way that $\Lambda_2^2/m << \Lambda_1$.
 The proper way to
implement this condition in dimensional regularization is again by expanding the fermion propagator in
pNRQED about the static propagator.

\medskip

Now we are in a position to prove that no pNRQED
diagram containing a loop contributes to the matching calculation. 

\medskip

Consider first the
two fermion Green function with an arbitrary number of ultrasoft legs.
 For potential terms to contribute we need at least a four
fermion Green function and hence we only have to care about ultrasoft photons. If we
input a momentum $\sim {\bf p}$ in the fermion line this momentum cannot flow out
through any external ultrasoft photon line (by definition of ultrasoft). Then it must
flow through the fermion line, which is a series of static propagators insensible to
the momentum flowing. Hence upon expanding about external fermion energies and
external energies and momenta of the ultrasoft photons there is no scale in any
integral and therefore any loop contribution vanishes. In fact, exactly the same
argumentation can be used for the NRQED calculation. Then we conclude that the terms
bilinear in fermions are exactly the same in NRQED and pNRQED. However, we have to keep
in mind that the latter (by definition) must be understood as containing ultrasoft
photons only. 

\medskip

Consider next the four fermion Green function in pNRQED containing several
potential terms but no US photon. Since no energy can flow through the potentials and
the static propagators are insensitive to the momentum, upon expanding about the US
external energy, the integrals over internal energies have no scale. However, these integrals
 have IR (pinch) singularities which are not regulated by standard DR.
 We shall take the additional prescription of putting
them to zero. Since the IR behavior of pNRQED and NRQED is the same, the same kind of integrals 
appear in the NRQED calculation. If we put them consistently to zero we obtain the correct
 potential terms, which play a r\^ole similar to the Wilson coefficients in the matching between QED
and NRQED. It is important to keep in mind that the
Wilson coefficients compensate for the different UV behavior of the effective theory (pNRQED)
with respect to that of the 'fundamental' theory (NRQED). Hence they are not sensitive to the
details of the IR behavior, which legitimates the prescription above.  
 Then any loop diagram in pNRQED with no US photons can be put to zero. This still holds if
 an arbitrary number of US photon
lines are included in the diagram. Indeed,
any potential line in the diagram now may also contain US momenta from the
photon lines. These, however, can be expanded about zero since they
are (by definition) much
smaller than the momentum transfer in the potential. Hence the
integrals over US photon 
energies and momenta contain
no scale (again upon expanding the US external energy in the fermion static
propagators) and can also be put to zero.

\medskip

In summary, we can directly identify the potential terms from a
 calculation in NRQED. We would like to stress again the similarity in the procedure
 with the matching between QED and NRQED as carried out in refs. \cite{Manohar, Match}.
 The potential terms in pNRQED play the r\^ole of Wilson
coefficients in the matching procedure.

\medskip

The four fermion terms appearing in the pNRQED lagrangian typically
 have the form\footnote{In principle (\ref{lpot}) could also depend on the total
 momentum ${\bf P}=-i \bfnabla_{{\bf X}}$,  with ${\bf X} = {{\bf x}_1
 + {\bf x}_2 \over 2}$, or on ultrasoft photons, but these effects can
 be neglected to the accuracy we are working at.}
\be
\label{lpot}
L^{pot} = -\int d^3{\bf x}_1 d^3{\bf x}_2 
\psi^{\dagger} (t,{\bf  x}_1)  \chi^{\dagger}_c (t,{\bf  x}_2)
V({\bf x}, {\bf p}, {\bfsigma}_1,{\bfsigma}_2)
\chi_c (t,{\bf  x}_2) \psi(t,{\bf  x}_1)
\ee
where ${\bf x} = {\bf x}_1 - {\bf x}_2$, ${\bf p} = -i
 \bfnabla_{{\bf x}}$ and ${\bf s}_1= {\bfsigma}_1/2$, ${\bf s}_2={\bfsigma}_2/2$ act on the fermion
 and antifermion, respectively (the spin fermion and antifermion
 indices are contracted with the potential indices, which are not explicitly
 displayed). $V({\bf x}, {\bf p}, {\bfsigma}_1,{\bfsigma}_2)$ may
 also be written as an expansion of the type
\be
V = V^{(0)}+V^{(1)}+V^{(2)}+V^{(3)}+ \cdots \,,
\ee
where $\langle V^{(n)} \rangle \sim m\al^n$. Our results are exact for
the four first terms of this expansion.

We obtain from the tree level diagrams of fig. 1 (${\tilde V}$ represents the
Fourier transform of $V$)

\bea
&&{\tilde V}^{(a)}_{tree} = -{ 4\pi\al \over {\bf k}^2}
\,,
\\
&&
{\tilde V}^{(b)}_{tree} = { \pi\al c_D \over m^2}
\,,
\\
&&
{\tilde V}^{(c)}_{tree} = - { i 2 \pi\al c_S \over m^2} { ({\bf p} \times
  {\bf k}) \cdot {\bf S} \over {\bf k}^2}
\,,
\\
&&
{\tilde V}^{(d)}_{tree} =- {16 \pi \alpha d_2 \over m^2}
\,,
\\
&&
{\tilde V}^{(e)}_{tree} = - { 4 \pi\al \over m^2} \left( { {\bf p}^2
  \over {\bf k}^2} - {({\bf p} \cdot {\bf k})^2 \over {\bf k}^4}
\right)
\,,
\\
&&
{\tilde V}^{(f)}_{tree} = - { i 4 \pi\al c_F \over m^2} { ({\bf p} \times
  {\bf k}) \cdot {\bf S} \over {\bf k}^2}
\,,
\\
&&
{\tilde V}^{(g)}_{tree} = { 4 \pi\al c_F^2 \over m^2} 
\left( {\bf s}_1  \cdot {\bf s}_2 - { {\bf s}_1 \cdot {\bf k} {\bf
      s}_2 \cdot {\bf k} \over {\bf k}^2} 
\right)
\,,
\\
&&
{\tilde V}^{(h)}_{tree} = (d_s+3d_v) -2d_v {\bf S}^2 
\,,
\eea
where ${\bf S} = {\bf s}_1 + {\bf s}_2 $, 
and for the one loop diagrams of fig. 2.
\be
{\tilde V}^{(a)}_{1loop} = { \al^2 \over m^2} \left(\log {{\bf k}^2
    \over \mu^2} - {8 \over 3}\log2 + {5 \over 3} \right)
\,,
\ee
\be
{\tilde V}^{(b,c)}_{1loop} = { 4\al^2 \over 3m^2} \left(\log {{\bf k}^2
    \over \mu^2} + 2 \log2 - 1 \right)
\,.
\ee
The $\mu$ dependence of fig. 2b,c is of IR origin and will eventually cancel with US
contributions. However, the  $\mu$ dependence of fig. 2a is of UV origin and cancels
exactly with the $\mu$ dependence of $d_s$. Recall that there is an additional $\mu$
dependence in $c_D$ which will also cancel against US contributions.
Upon Fourier transforming and putting together the above results we
obtain 
\be
V = V^{(0)} + \delta V \,,
\ee
where
\be
V^{(0)} = - {\al \over \vert {\bf x} \vert}
\ee
and
\bea
&&\delta V = - { \al \over 2 m^2} { 1 \over |{\bf x}|}
       \left( {\bf p}^2 + { 1 \over {\bf x}^2} {\bf x} \cdot
                 ({\bf x} \cdot {\bf p}){\bf p} \right)
\\
&&
\nonumber
+ { \delta^{(3)}({\bf x}) \over m^2}
       \left( \pi \al \left(c_D -2c_F^2 \right) +d_{s}+3d_{v} -16\pi
         \alpha d_2 +{\al^2 \over 3} - {7 \al^2 \over 3}\log \mu^2 \right) -
       { 7 \al^2 \over 6 \pi m^2} {\rm reg} {1 \over |{\bf x}|^3}
\\
&&
\nonumber
+ { \delta^{(3)}({\bf x}) \over m^2} {\bf S}^2
       \left( \pi \al { 4 \over 3}c_F^2 -2 d_{v} \right)
+ { \al \over 4 m^2} { 1 \over |{\bf x}|^3} {\bf L} \cdot {\bf S}
       \left( 2c_S+4c_F \right)
+ { \al c_F^2 \over 4 m^2} { 1 \over |{\bf x}|^3}
            S_{12} ({\hat {\bf x}})
\,,
\eea
where $S_{12} ({\hat {\bf x}}) = (-{\bfsigma}_1 \cdot {\bfsigma}_2 +3
       {\bfsigma}_1 \cdot {\hat {\bf x}} \, {\bfsigma}_2 \cdot {\hat {\bf
       x}})$ and (see \cite{ynd} for more details on the Fourier transform)
\be
-{1 \over 4\pi} { \rm reg} {1 \over |{\bf x}|^3} 
=  \int { d^3{\bf k} \over (2\pi)^3} e^{i{\bf k} \cdot {\bf x}}\log k 
\, .
\ee 
The pNRQED lagrangian now reads
\bea
\label{lpq}
\nn
&& L_{pNRQED}= \int d^3{\bf x}\Biggl(\psi^{\dagger} \Biggl\{ i D^0 + \,
 {{\bf D}^2\over 2 m} + {{\bf D}^4\over8 m^3} \Biggr\} \psi
 + (\chi_c, e \rightarrow -e)
  - {1\over 4} F_{\mu \nu} F^{\mu \nu} \Biggr) \\
&&\quad -\int d^3{\bf x}_1 d^3{\bf x}_2 
\psi^{\dagger} (t,{\bf  x}_1)  \chi^{\dagger}_c (t,{\bf  x}_2)
V({\bf x}, {\bf p}, {\bfsigma}_1,{\bfsigma}_2)
\chi_c (t,{\bf  x}_2) \psi(t,{\bf  x}_1)  \,,\eea
where the photons are ultrasoft.

In order to make explicit the size of each term in (\ref{lpq}) it is convenient to project pNRQED
to the one electron one positron subspace (this can be easily done at
the hamiltonian level). This subspace is spanned by
\be
 \int d^3{\bf x}_1 d^3{\bf x}_2 \varphi ({\bf x}_1,{\bf x}_2 ) \psi^{\dagger}({\bf x}_1)
 \chi^{\dagger}_c({\bf x}_2) \vert 0 \rangle
\,,
\ee
where $\vert 0 \rangle$ is the subspace of the Fock space containing zero electrons and
positrons but an arbitrary number of ultrasoft photons.

\medskip
Then the dynamics of the wave function field is described by the lagrangian
\bea
\label{6}
&&L_{pNRQED}= 
\int d^3{\bf x}_1 d^3{\bf x}_2 \varphi^{\dagger} ({\bf x}_1,{\bf x}_2, t)
\Biggl(i\partial_0 + {\bfnabla^2_{\bf x}\over  m}+ {\bfnabla^2_{\bf
    X}\over 4 m} +{\bfnabla^4_{\bf x}\over 4 m^3} -ex^i\partial_i A_{0} (t,{\bf X})
\nn
\\
&&
\quad \quad 
 -2ie{{\bf A} (t,{\bf X}) \cdot \bfnabla_{\bf x} \over  m}
- V({\bf x}, {\bf p}, {\bfsigma}_1,{\bfsigma}_2) \Biggr)\varphi ({\bf x}_1,{\bf x}_2,t)
- \int d^3{\bf x} {1\over 4} F_{\mu \nu} F^{\mu \nu}
\,,
\eea
where we have made precise that the remaining photon fields are ultrasoft by multipole
expanding them about the center of mass\footnote{As expected for a
  chargeless particle the covariant derivatives for ${\bf P}$ are the
  ordinary ones.}. Furthermore, gauge invariance at any order in
the multipole expansion can be made explicit by introducing
\be
\varphi ({\bf x_1} ,{\bf x_2} , t)= P\bigl[e^{ie\int_{{\bf x_2}}^{{\bf
x_1}} {\bf A} \cdot d {\bf x}} \bigr]S({\bf x}, {\bf X}, t)
\,.
\ee
Then, the gauge transformations of the above wave function fields are
\be
\nonumber
\varphi ({\bf x_1} ,{\bf x_2} , t)\rightarrow g({\bf x_1} ,t)\psi ({\bf
x_1} ,{\bf x_2} , t)
g^{-1}({\bf x_2} ,t )
\,,
\ee
$$
S ({\bf x}, {\bf X}, t)\rightarrow S ({\bf x} ,{\bf X} , t)
\,.
$$
We finally obtain 
\bea
\label{lpnrqed}
&&L_{pNRQED} =
\int d^3{\bf x} d^3{\bf X} dt S^{\dagger}({\bf x}, {\bf X}, t)
                \Biggl\{
i\partial_0 - { {\bf p}^2 \over m} + { {\bf p}^4 \over 4m^3} - { {\bf P}^2 \over4 m}
\\
&&
\nonumber
- V ({\bf x}, {\bf p}, {\bfsigma}_1,{\bfsigma}_2) + e {\bf x} \cdot {\bf E} ({\bf X},t)
\Biggr\}
S ({\bf x}, {\bf X}, t)- \int d^3{\bf x} {1\over 4} F_{\mu \nu} F^{\mu \nu}
\,,
\eea
which is explicitly gauge invariant. 
 Moreover the size
of each term is unique and can be evaluated as follows. Each relative momentum 
$\partial_{\bf x}$,
and inverse relative coordinate $\vert
{\bf x}\vert^{-1}$ have a size $\sim m\alpha$. Each US photon field, 
derivatives acting
on it, the time derivative and the center of mass momentum 
$\partial_{\bf X}$ (in the
rest frame, when entering in recoil corrections due to the virtual emission of US
 photons) 
on the wave function field have a size $\sim m\alpha^2$.
Formula (2.23) has already been presented in \cite{Mont} (except for a numerical factor in the
potential which was
left arbitrary).
 We shall use it to
calculate the spectrum at $O(m\alpha^5)$ in the next section.

\medskip

The gauge independence of the matching calculation is checked at
$O(m\al^4)$ in Appendix A.
The pNRQED Lagrangian for the unequal mass case
can be built with no
further difficulty. The result is displayed in Appendix B.

\bigskip

\section{ Bound state calculation in pNRQED} 
\indent

In order to find the corrections to the bound state energy of a state with principal
quantum number $n$ we consider the following Green function (we will
follow a procedure and notation similar to ref. \cite{Lamb})
\be
\Pi(q,{\bf x}):=\int d x^0 d{\bf X} e^{iqx^0}<T\{ \varphi (0)
\varphi^{\dagger}({\bf x}, {\bf X}, t) \}>
\ee
when $q \rightarrow E_{n}$, where $E_n$ is the energy of the leading
hamiltonian 
\be
\hat h_0=-{\bfnabla^2\over m}-{\al\over\vert {\bf x} \vert }
\,.
\ee
The integral over ${\bf X}$ fixes the center of mass momentum ${\bf
  P}$ to zero. We write
\be
\Pi(q,{\bf x})={A_n +\d A_n \over q-(E_n+\d E_n )}\sim {A_n +\d A_n \over q-E_n}+
{A_n \over q-E_n}\d E_n{1 \over q-E_{n}}
\,.
\ee 

\medskip

The contribution to $\delta E_n$ coming from the correction to the
potential and the kinetic energy read
\bea
&&\d^{V} E_{n} = \langle nlj\vert \delta V \vert nlj \rangle = {m
  \al^4 \over 8} \Biggl\{ - {3 \over n^3(2l+1)} + {1 \over n^4}
\nn
\\ 
\nn
&&
-{2 \al \over 3 \pi}{ \delta_{l0} \over n^3}\left(- \log {m \over \mu}
  +7\log {\mu n \over m \al} -6 \log 2 + {17 \over 5} -7 \left(\sum^n_{k=1} {1 \over
  k} + {n-1 \over 2n} \right)\right)
\\ 
&& -{7 \al \over 3 \pi}{1- \delta_{l0} \over n^3} {1 \over l(l+1)(2l+1)} 
\\
\nn
&&
+{14 \over 3}{ \delta_{l0}\delta_{s1} \over n^3}\left\{1+{3 \al \over 7
    \pi} \left(- {32 \over 9} -2 \log 2 \right)\right\} 
+
 {(1- \delta_{l0}) \delta_{s1} \over l(2l+1)(l+1)n^3}\,C_{j,l}
\Biggr\} 
\,,
\eea
\be
\d^{K} E_{n} =-{1 \over 4m^3} \langle nlj\vert \bfnabla^4 \vert nlj
  \rangle =  {m \al^4 \over 8} {3(l+1/2)-4\,n \over 4\,n^4(2\,l+1)} \,,
\ee
where 
\begin{eqnarray}
 C_{j,l} = \,\left\{
\begin{array}{ll}
 \displaystyle{
-{l+1 \over 2\,l-1} \left(2(3l-1)+{\al \over \pi}(4\,l-1) \right)}
 &\ \ , \, j=l-1\;,
\\
 \displaystyle{
-2-{\al \over \pi}}
 &\ \ ,\, j=l\;,
\\
\displaystyle{
{l \over 2\,l+3} \left(2(3l+4)+{\al \over \pi}(4\,l+5) \right)}
 &\ \ ,\, j=l+1\;.
\end{array}
\right.
\end{eqnarray}

There is also a contribution from a virtual exchange of an ultrasoft photon
corresponding to the diagram in fig. 3, which has already been evaluated in dimensional regularization for the
Hydrogen-like atom \cite{Lamb} (here
the calculation is identical but using the reduced mass). Notice that
the $\overline{MS}$ scheme has to be used in the calculation. Since
(\ref{lpnrqed}) is gauge invariant we can use any gauge
to calculate this contribution. Still the Coulomb gauge continues to
be advantageous, since in this gauge $A_0$ can only contribute to tadpoles which can be safely put to
zero in dimensional regularization.
This contribution reads
\bea
&&\d^{US} E_{n} = - {8\over 3}{\al\over \pi} \sum_{m}
\vert
\langle n\vert {{\bf p} \over m} \vert m \rangle
\vert^2
\left( E_{n} -E_{m}\right) \left( \log{\mu \over\vert  E_{n}
    -E_{m}\vert} -\log 2 +{5 \over 6} \right)
\\
\nn
&&
\quad\quad
= -{m \al^5 \over 3 \pi n^3} \left(\d_{l0}\left[\log {\langle
      E_{n,l}\rangle^2 \over \mu^2} -{5 \over 3}\right] + (1-\d_{l0})
  \log {4 \langle E_{n,l}\rangle^2 \over m^2\al^4} \right)
\,,
\eea
where the last equation implicitly defines $\langle E_{n,l}
\rangle$. Once we add all these contributions the final result reads
\begin{eqnarray}
 &&
\delta E_{n,l,j} = \d^{V} E_{n} + \d^{K} E_{n} + \d^{US} E_{n} = {m
  \al^4 \over 8} \Biggl\{ - {4 \over n^3(2l+1)} + {11 \over 8n^4}
\nn
\\ 
\nn
&&
-{2 \al \over 3 \pi}{ \delta_{l0} \over n^3}\left( 9\log\al +7\log n +
8\log R(n,l) -14 \log 2 - {49 \over 15} -7 \left(\sum^n_{k=1} {1 \over
  k} + {n-1 \over 2n} \right)\right)
\\ 
&& -{16 \al \over 3 \pi}{1- \delta_{l0} \over n^3}\left( \log R(n,l)
  + {7 \over 16} {1 \over l(l+1)(2l+1)} \right) 
\\
\nn
&&
+{14 \over 3}{ \delta_{l0}\delta_{s1} \over n^3}\left\{1+{3 \al \over 7
    \pi} \left(- {32 \over 9} -2 \log 2 \right)\right\} 
+
 {(1- \delta_{l0}) \delta_{s1} \over l(2l+1)(l+1)n^3}\,C_{j,l}
\Biggr\}
\,,
\end{eqnarray}
where $\log R(n,l)= \log{2 \langle E_{n,l}\rangle \over m\al^2}$ is
called the Bethe logarithm. For the $O(m\al^5)$ contribution we find
agreement with ref. \cite{positronium} for the spin independent piece
and for the $\d_{l0}\d_{s1}$
piece, while for the $(1-\delta_{l0})\delta_{s1}$ piece we find
agreement with ref. \cite{Labelle2} (this last piece could also be
obtained from results of ref. \cite{positronium}). 

\medskip

We can also easily obtain the full decay width at lowest order. It reads
\be
\Gamma_n= {m \al^5 \over 2 n^3}\d_{10}(1-\d_{s1})+\sum_{m<n}{16\over 3}\al
\vert
\langle n\vert {{\bf p} \over m} \vert m \rangle \vert^2(E_n-E_m)
\,.
\ee

\bigskip

\section{Discussion and Conclusions}
\indent

We have seen that pNRQED correctly reproduces the positronium spectrum at
$O(m\alpha^5)$. This is a non-trivial check of the ideas behind this EFT
since at this order all regions of momenta (hard, soft and ultrasoft) 
 contribute to the
energy.

\medskip

We would like to emphasize that the procedure we propose for higher order
corrections to the positronium (and other QED bound states) is totally
systematic. It uses two EFTs, namely NRQED and pNRQED. Both the matching 
from QED to NRQED and from NRQED to pNRQED can be done order by order in
$1/m$ and $\alpha$, and static propagators for the fermions can be used.
This together with the use of dimensional regularization simplifies a lot the calculations. The
actual bound state calculation is done at the level of pNRQED for the wave
function field and is very similar to a standard quantum mechanical
calculation. The only difference being that the wave function field couples
to US photons in a field theoretical fashion. 

\medskip

We believe that the clarity and simplicity of this formalism  
will allow to carry out higher order bound state calculations in QED very
efficiently. In order to illustrate this point let us pose ourselves the 
calculation of the positronium spectrum at $O(m\alpha^6)$ and see the extra calculations 
required in order to obtain the pNRQED lagrangian at this order. Clearly all
contributions that we obtain at $O(m\alpha^5)$ which are multiplied by a
Wilson coefficient, will give a contribution at $O(m\alpha^6)$ by just
calculating the Wilson coefficient to next order in $\alpha$. This requires
the matching from QED to NRQED at two loops. The relativistic correction to the kinetic
energy 
$O(1/m^5)$ in NRQED should be kept. Terms $O(1/m^4)$ in the
NRQED lagrangian would now contribute to the potential but they do so only
 at tree level. Hence
their Wilson coefficients are only necessary at tree level. Those for
the bilinear terms may be obtained from \cite{Balk} whereas those for
the four fermion operators from \cite{Lepage2}.
Terms $O(1/m^3)$ also contribute at tree level and may contribute at one loop.
In either case the Wilson coefficients are only needed at tree level
 which
are known \cite{Manohar}. In addition $O(1/m^2)$ terms in the NRQED
lagrangian would now contribute to the potential at one loop and $O(1/m)$
terms at two loops. It is also easy to see by inspecting the next order
terms of the multipole expansion in the pNRQED
lagrangian that, due to angular momentum conservation, no contribution of US
photons arises at $O(m\alpha^6)$. There would only be a new qualitative
feature, namely that time derivatives multiplying potential terms would
arise (for instance, from the expansion of the energy in the one
transverse photon exchange at tree level). This time derivatives can be
disposed of by using the equations of motion in pNRQED (now with potential
terms included) as it has been
done in Appendix A, according to the philosophy of ref. \cite{Fearing}
(see also \cite{Balzereit}). This
calculation would produce an independent check of the existing results
 obtained very recently in ref. \cite{ma6}.

\medskip
   
 It is important to be able to calculate systematically higher orders 
in QED as a test to the Standard Model in a sector where QCD does not
play any 
relevant role. Any signal of new physics here should be much clearer
 than in 
the hadronic sector as, for instance, in the orthopositronium decay 
where there
seem to be some difficulties to explain the data (see \cite{Lepage3}
and references there in)\footnote{Only one experiment \cite{japon}
seems to be compatible with theoretical predictions.}.

\medskip

Finally, we would like to stress that the idea of separating the 
calculation of the binding energy (or any other observable) of
a non-relativistic bound state system in three stages, namely (i)
integrating out the hard scale, (ii) integrating out the soft scale and
(iii) calculating the bound state energy when only ultrasoft
 degrees of freedom
are present, is neither confined to positronium nor to the QED realm.
Indeed, we have already shown in ref. \cite{Lamb} that it can be 
applied to
Hydrogen-like atoms, and it should be easy to work out pNRQED for
 muonium,
di-muonium, and other two body QED bound states. In particular, pionium, a
QED bound state which, however, decays strongly, has
received considerable attention lately \cite{pionium}. Its decay width 
turns out to be
proportional to the pion scattering length, which is an essential input to
fix the parameters of the chiral lagrangian \cite{GL}. In order to extract
the scattering length neatly from the experimental data a good control on
the electromagnetic corrections is necessary. pNRQED for pionium can
definitely help in that goal. Beyond QED, heavy
quarkonium systems also form non-relativistic bound states. We have
already proposed that potential NRQCD (pNRQCD), an EFT for NRQCD 
analogous to pNRQED for NRQED, should be useful to study these systems
\cite{Mont}. The techniques presented here may also help in the
understanding of nucleon-nucleon bound states from the heavy baryon 
chiral lagrangian, which have also received quite some attention during
the last years \cite{nuclear}.   

\medskip

{\bf Acknowledgments}
\medskip

Financial support from CICYT, contract AEN95-0590 and
 from CIRIT, contract GRQ93-1047 is
acknowledged. A.P. acknowledges financial support from the TMR network under
contract FMRX-CT96-0008.

\bigskip

\appendix

\Appendix
{:  $O(m\al^4)$ matching in the Feynman gauge}
\indent

In this Appendix we check that the pNRQED lagrangian is gauge independent
at $O(m\al^4)$ once it is written in the standard form. By the standard form we 
understand that all time derivatives in higher order terms are disposed of by the use of equations of
motion, or alternatively by local field redefinitions
\cite{Manohar,Fearing} (see also \cite{Balzereit}). 
In order to do so the matching calculation is redone
in the Feynman gauge. 

\medskip

The main difference between the Coulomb gauge and the Feynman gauge, as far as the matching of the four
point Green function is concerned, is that in the former loop diagrams involving $A_0$ only  
can be set to zero in NRQED
because there is no scale for the integration over the energy, whereas in the latter they must be kept because
the poles of the $A_0$ propagator now relate energy and momentum.

\medskip

The counting in section 2 implies now that the following extra diagrams must be considered: (i) the one
loop diagrams of fig. 4 ($O(m\alpha^3)$), (ii) the same diagrams with a ${\bf p}^2/2m$ insertion in either
fermionic line ($O(m\alpha^4)$), (iii) the same diagrams with external energy insertions arising from
the expansion of the propagators about zero external energy ($O(m\alpha^4)$), and (iv) two loop diagrams
involving $A_0$ propagators only ($O(m\alpha^4)$).

\medskip

It is easy to see that diagrams (i) cancel each other. Diagrams (ii) and (iii) vanish individually, essentially
because they have an odd number of static propagators. Diagrams (iv) have already been seen to cancel in
ref. \cite{Peter}. Then we are left with the same diagrams we had in the Coulomb gauge (all diagrams
in fig. 1, except
(d) which is $O(m\alpha^5)$),
 but now they
must be calculated in the Feynman gauge.

In fact, all the diagrams
give the same result except fig. 1a and fig. 1e. The latter now reads
\be
\label{feyntr}
{\tilde V}^{(e)}_{tree}({\rm Feynman}) = {\tilde V}^{(e)}_{tree}({\rm
  Coulomb}) - { \pi\al \over m^2} { \left( {\bf p}^2 - {\bf
      p^{\prime}}^2 \right)^2 \over {\bf k}^4}
\,.
\ee
Fig 1a  now receives a contribution
 due to the expansion of the external energies about zero in the $A_0$ propagator, which we depicted in
  (fig. 5).
 It reads
\be
\label{feynlong}
{\tilde V}^{new}_{tree}({\rm Feynman})= - 4\pi\al \left( {k^0 \over
    {\bf k}^2} \right)^2
\,,
\ee
where $k^0=E_1 -E_1^{\prime}$. $E_1$ and $E_1^{\prime}$ give rise
to time derivatives in the pNRQED lagrangian. 
\bea
\label{lagfeyn}
&&L^{new}({\rm Feynman}) = -\int d^3{\bf x}_1 d^3{\bf x}_2 
\partial_0^2 \left( \psi^{\dagger} \psi(t,{\bf  x}_1) \right)
\int { d {\bf k} \over (2\pi)^3} e^{i{\bf k}\cdot{\bf x}}{1 \over {\bf k}^4}
\chi^{\dagger}_c \chi_c (t,{\bf  x}_2)
\nn
\\
&&
\quad\quad
= \int d^3{\bf x}_1 d^3{\bf x}_2 
\partial_0 \left( \psi^{\dagger} \psi(t,{\bf  x}_1) \right)
\int { d {\bf k} \over (2\pi)^3} e^{i{\bf k}\cdot{\bf x}}{1 \over {\bf k}^4}
\partial_0 \left(\chi^{\dagger}_c \chi_c (t,{\bf  x}_2)\right) 
\eea
We can get rid of these time derivatives by
using the equations of motion. Notice however that now potential terms
enter in the equations of motion. 
Their explicit inclusion can be avoided in this case by using the continuity equation in
the last equality of (\ref{lagfeyn})
\be
{\partial \rho \over \partial t} +\bfnabla \cdot {\bf j}=0
\,,
\ee
where $\rho=\psi^{\dagger}\psi$ and
\be
{\bf j}=- {i \over 2m} \left[\psi^{\dagger}\bfnabla \psi - (\bfnabla
  \psi^{\dagger}) \psi \right]
\,.
\ee
(\ref{feynlong}) can now be written as 
\be
{\tilde V}^{new}_{tree}({\rm Feynman})= + { \pi\al \over m^2} { \left( {\bf p}^2 - {\bf
      p^{\prime}}^2 \right)^2 \over {\bf k}^4}
\,,
\ee
which just cancels the extra contribution in (\ref{feyntr}). We have then proved that the pNRQED
lagrangian written in the standard form (i.e. with no time derivatives) at $O(m\alpha^4)$ is exactly the
 same
in the Coulomb and Feynman gauges.  
Notice that it has been crucial to write the time derivatives in a
    symmetric fashion in order to use the continuity equation. The na\"\i ve use of the on-shell
condition $k^0= {\bf p}^2/2m - {\bf
      p^{\prime}}^2/2m$ in (A.2) leads to incorrect results. 

\bigskip
 
\Appendix{: pNRQED lagrangian for the unequal mass case}
\indent

Here we display the lagrangian relevant for the calculation of the
mass to $O(m\al^5)$ for the unequal mass case (we assume $m_1$, $m_2
>> {\bf p} >> E$). The charge of each
particle has opposite sign.
\bea
&&L_{pNRQED} =
\int d^3{\bf x} d^3{\bf X} dt S^{\dagger}({\bf x}, {\bf X}, t)
                \Biggl\{
i\partial_0 - { {\bf p}^2 \over \mu_{12}} + { {\bf p}^4 \over 8m_1^3}+ 
{ {\bf p}^4 \over 8m_2^3} - { {\bf P}^2 \over2M}
\\
&&
\nonumber
- V ({\bf x}, {\bf p}, {\bfsigma}_1,{\bfsigma}_2) + e {\bf x} \cdot {\bf E} ({\bf X},t)
\Biggr\}
S ({\bf x}, {\bf X}, t)- \int d^3{\bf x} {1\over 4} F_{\mu \nu} F^{\mu \nu}
\,,
\eea
where $M=m_1+m_2$, $\mu_{12}= {m_1m_2 \over m_1+m_2}$, ${\bf x}$ and ${\bf X}$, and
${\bf p}$ and ${\bf P}$ are the relative and center of mass coordinate and momentum
respectively.
 The potential
now reads
\bea
&& V =  - {\al \over \vert {\bf x} \vert} 
 - { \al \over 2 m_1m_2} { 1 \over |{\bf x}|}
       \left( {\bf p}^2 + { 1 \over {\bf x}^2} {\bf x} \cdot
                 ({\bf x} \cdot {\bf p}){\bf p} \right)
\\
&&
\nonumber
+ { \delta^{(3)}({\bf x}) \over m_1 m_2}
       \Biggl[ \pi \al \left({c_D^{(2)}m_1^2+c_D^{(1)}m_2^2 \over
             2m_1m_2}-2c_F^2 \right) +d_{s}+3d_{v} -16\pi
      \alpha d_2 \left({m_1^2+m_2^2 \over m_1m_2} \right)+{\al^2 \over 3} 
\\
&&
\nn
- {7 \al^2 \over 3}\log \mu^2 \Biggr] -
       { 7 \al^2 \over 6 \pi m_1m_2} {\rm reg} {1 \over {\vert {\bf x}
           \vert}^3}
+ { \delta^{(3)}({\bf x}) \over m_1m_2} {\bf S}^2
       \left( \pi \al { 4 \over 3}c_F^2 -2 d_{v} \right)
\\
&&
\nonumber
+ { \al c_F \over m_1m_2} { 1 \over |{\bf x}|^3} {\bf L} \cdot {\bf S}
+ { \al c_S \over 2 m_1m_2} { 1 \over |{\bf x}|^3} {\bf L} \cdot
       \left( {{\bf s}_2 m_1^2+{\bf s}_1m_2^2 \over
             m_1m_2} \right)
+ { \al c_F^2 \over4 m_1m_2} { 1 \over |{\bf x}|^3}
            S_{12} ({\hat {\bf x}})
\,,
\eea
where 
\be
c_D^{(i)}= 1+{\alpha\over \pi} \left(
{4\over 3} \log {m_i^2 \over \mu^2} \right) 
\ee  
and now $d_s$ and $d_v$ read (see \cite{Mont})
\be
d_{s}=
  - {\al^2 \over m_1^2-m^2_2}
\left\{m_1^2\left(  \log{m^2_2 \over  \mu^2}
                   + {1 \over 3} \right)
       -
       m^2_2\left(  \log{m^2_1 \over  \mu^2}
                   + {1 \over 3} \right)
\right\}
\,,\ee
\be
d_{v}=
   {\al^2 \over m_1^2-m^2_2}
m_1 m_2\log{m^2_1 \over m^2_2}
\,.\ee

\bigskip

The lagrangian (B.2) must be corrected if there are charged particles
of masses $m_{i}$, $i=3,4..$,
similar or smaller than $m:=max\{m_1 ,m_2\}$. Each
particle of mass $m_{i}$  such that $m 
{\ \lower-1.2pt\vbox{\hbox{\rlap{$>$}\lower5pt\vbox{\hbox{$\sim$}}}}\
  } m_i  >> \mu_{12}\alpha $ gives an extra contribution
$1/m_{i}^2$ multiplying to $d_2$ in (B.2). Each particle of mass
$\mu_{12}\alpha 
{\ \lower-1.2pt\vbox{\hbox{\rlap{$>$}\lower5pt\vbox{\hbox{$\sim$}}}}\
  } m_i >> \mu_{12}\alpha^2 $ gives extra
non-trivial contributions to the potential \cite{pionium}.



\vfill
\eject

{\bf Caption1.} The nonzero relevant diagrams for the matching at tree level
       in the Coulomb gauge. The dashed and zigzag lines represent the
       $A_0$ and ${\bf A}$ fields respectively, while the continuous
       lines represent the fermion and antifermion fields. The first diagram
       is the Coulomb potential. For the $A_0$ the circle is the vertex
       proportional to $c_D$, the square to $c_S$ (spin dependent) and
       the dashed dot to $d_2$ (the vacuum polarization), while for
       ${\bf A}$ the square is the vertex proportional to $c_F$ and
       the other vertex appear from the covariant derivative in the
       kinetic term. The last diagram is proportional to $d_s$ and
       $d_v$. The symmetric diagrams are not displayed.

\bigskip

{\bf Caption2.} The nonzero relevant diagrams for the matching at one loop
       in the Coulomb gauge. The dashed and zigzag lines represent the
       $A_0$ and ${\bf A}$ fields respectively, while the continuous
       lines represent the fermion and antifermion. The interactions
       for ${\bf A}$ are the ones which appear from the covariant space
       derivatives in the kinetic term,
       while for $A_0$ comes from the covariant time derivative. The
       symmetric diagrams are not displayed.

\bigskip

{\bf Caption3.} The thick line and wavy line are the positronium and the
  transverse ultrasoft photon propagators respectively.

\bigskip

{\bf Caption4.} The nonzero relevant diagrams for the matching at one loop
       in the Feynman gauge setting $E$ to zero and with no $1/m$ insertions.

\bigskip

{\bf Caption5.} Correction to the $A_0$ propagator due to energy insertions
  in the Feynman gauge.


\begin{thebibliography}{99}

\bibitem{Lepage} W.E. Caswell and G.P. Lepage, {\it Phys. Lett.} {\bf
B167} (1986) 437. 

\bibitem{Kinoshita} T. Kinoshita and M. Nio,
{\it Phys. Rev.} {\bf D53} (1996) 4909.

\bibitem{Labelle2} P. Labelle, S.M. Zebarjad and C.P. Burgess,
{\it Phys. Rev.} {\bf D56} (1997) 8053.

\bibitem{LH} A.H. Hoang, P. Labelle and S.M. Zebarjad, 
{\it Phys.Rev.Lett.} {\bf 79} (1997) 3387.

\bibitem{BS} H.A. Bethe and E.E. Salpeter,
{\it Phys. Rev.} {\bf 82} (1951) 309.

\bibitem{Labelle}
P. Labelle, {\it Effective field theories for QED bound states:
extending Nonrelativistic QED to study retardation effects}, McGill/96-33,
hep-ph/9608491.

\bibitem{LM} M. Luke and A.V. Manohar, {\it Phys. Rev.} {\bf D55}
  (1997) 4129. 

\bibitem{Grin} Grinstein and I.Z. Rothstein, {\it Phys. Rev.} {\bf
    D57} (1998) 78.

\bibitem{Sav} M. Luke and M.J. Savage, {\it Phys. Rev.} {\bf D57} (1998) 413.

\bibitem{Mont} A. Pineda and J. Soto, {\it Nucl. Phys. B (Proc. Suppl.)} {\bf 64}
(1998) 428.

\bibitem{Lamb} A. Pineda and J. Soto, 
{\it Phys. Lett.} {\bf B420} (1998) 391.

\bibitem{Beneke} M. Beneke and V.A. Smirnov, {\it Nucl. Phys.} {\bf
B522} (1998) 321. 
 
\bibitem{Harald} H.W. Griesshammer, {\it The Soft Regime in NRQCD},
  NT@UW-98-12, hep-ph/9804251.

\bibitem{Manohar} A.V. Manohar, {\it Phys.. Rev.} {\bf D56} (1997) 230.

\bibitem{Match} A. Pineda and J. Soto,
{\it Matching at one loop for the four-quark operators in NRQCD}, 
hep-ph/9802365, to be published in {\it Phys. Rev. D}.

\bibitem{Fearing} S. Scherer and H.W. Fearing, 
{\it Phys. Rev.} {\bf D52} (1995) 6445.

\bibitem{Balzereit} C. Balzereit, {\it Renormalizing Heavy Quark
    Effective Theory at O($1/m_Q^3$)}, hep-ph/9801436.

\bibitem{ynd} S. Titard and F.J. Yndur{\'a}in, 
{\it Phys. Rev.} {\bf D49} (1994) 6007.

\bibitem{positronium} S.N. Gupta, W.W. Repko and C.J. Suchyta III,
{\it Phys. Rev.} {\bf D40} (1989) 4100.

\bibitem{Balk} S. Balk, A. Ilakovac, J. C. K\"orner and D. Pirjol,
  Ahrenshoop Symp. 1993: 315-326 (QCD161:S973:1993).

\bibitem{Lepage2} G.T. Bodwin, E. Braaten and G.P. Lepage, 
{\it Phys. Rev.} {\bf D51} (1995) 1125, Erratum {\it ibid} {\bf D55} 
(1997) 5853.

\bibitem{ma6} K. Pachucki and S.G. Karshenboim, 
{\it Phys. Rev. Lett.} {\bf 80} (1998) 2101.

\bibitem{Lepage3} P. Labelle, G.P. Lepage and U. Magnea,
{\it Phys. Rev. Lett.} {\bf 72} (1994) 2006.

\bibitem{japon} S. Asai, S. Orita and N. Shinohara,
{\it Phys. Lett.} {\bf B357} (1995) 475.

\bibitem{pionium} P. Labelle and K. Buckley,
{\it A new order $\alpha$ correction to the decay rate of pionium}, 
hep-ph/9804201.

\bibitem{GL} J. Gasser and H. Leutwyler, {\it Ann. Phys.} {\bf(N.Y.)158}
(1984) 142; {\it Nucl. Phys.} {\bf B250} (1985) 465.

\bibitem{nuclear} S. Weinberg,
{\it Phys. Lett.} {\bf B251} (1990) 288;
{\it Nucl. Phys.} {\bf B363} (1992) 3.
C. Ord{\'o}{\~n}ez, L. Ray and U. van Kolck,
{\it Phys. Rev. Lett.} {\bf 72} (1994) 1982;
{\it Phys. Rev.} {\bf C53} (1996) 2086.
D.B. Kaplan, M.J. Savage and M.B. Wise, nucl-th/9802075.

\bibitem{Peter}M. Peter,
{\it Phys. Rev. Lett.} {\bf 78} (1997) 602; 
{\it Nucl. Phys.} {\bf B501} (1997) 471.

\end{thebibliography}
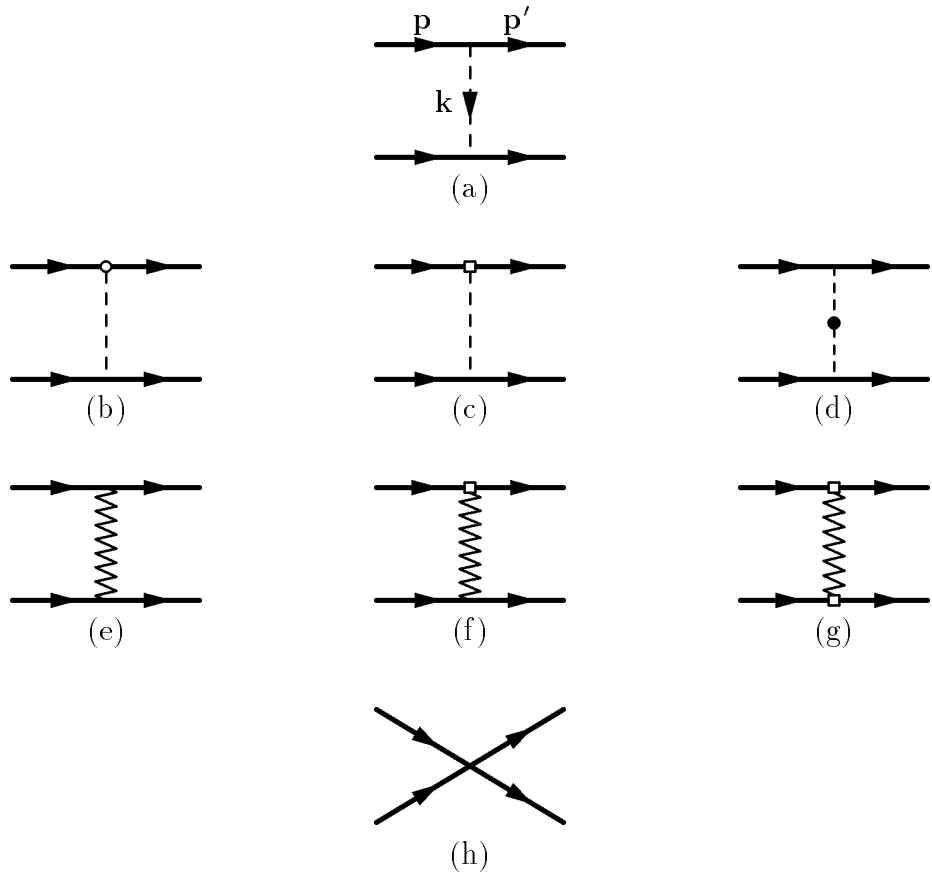
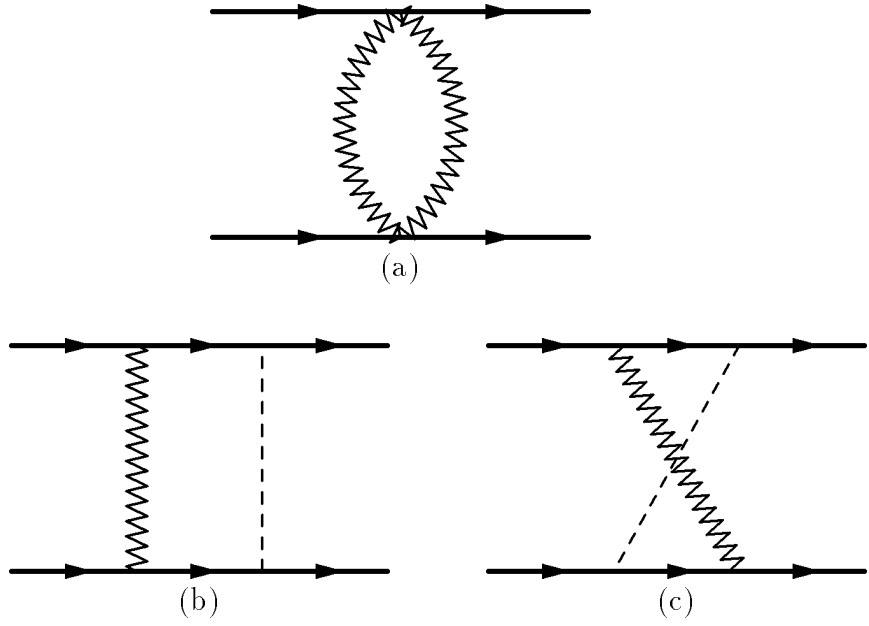
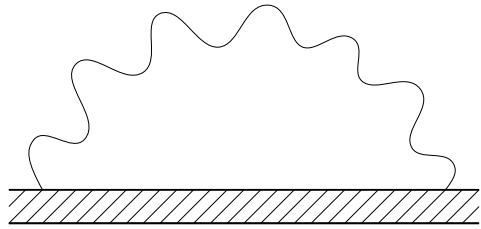
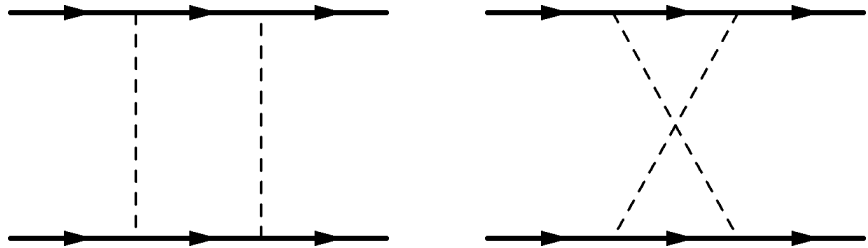
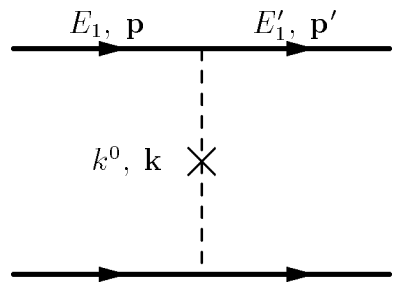
\end{document}